\documentclass[prb,floatfix,twocolumn,showpacs,amsmath,amssymb]{revtex4}
\usepackage{graphicx}
\usepackage{epstopdf}
\usepackage{epsfig}
\usepackage{dcolumn}
\usepackage{bm}
\usepackage[colorlinks=true,linkcolor=blue,citecolor=blue]{hyperref}

\begin{document}
\title{Variational Monte Carlo study of chiral spin liquid in the extended Heisenberg model on the Kagome lattice}
\author{Wen-Jun Hu,$^1$ Wei Zhu,$^1$ Yi Zhang,$^2$ Shoushu Gong,$^1$ Federico Becca,$^3$ and D. N. Sheng$^1$}
\affiliation{
$^1$ Department of Physics and Astronomy, California State University, Northridge, California 91330, USA \\
$^2$ Department of Physics, Stanford University, Stanford, California 94305, USA \\
$^3$ Democritos National Simulation Center, Istituto Officina dei Materiali del CNR, and SISSA-International School for Advanced Studies, Via Bonomea 265, I-34136 Trieste, Italy
}

\begin{abstract}
We investigate the extended Heisenberg model on the Kagome lattice by using Gutzwiller 
projected fermionic states and the variational Monte Carlo technique. In particular, when
both second- and third-neighbor super-exchanges are considered, we find that a gapped spin 
liquid described by non-trivial magnetic fluxes and long-range chiral-chiral correlations is 
energetically favored compared to the gapless U(1) Dirac state. Furthermore, the topological 
Chern number, obtained by integrating the Berry curvature, and the degeneracy of the ground state,
by constructing linearly independent states, lead us to identify this flux state as the chiral 
spin liquid with $C=1/2$ fractionalized Chern number.
\end{abstract}

\pacs{75.10.Jm, 75.10.Kt, 75.40.Mg, 75.50.Ee}

\maketitle

{\it Introduction --} 
Quantum spin liquids (QSL) are exotic phases of strongly correlated spin systems, which do
not possess any local order even at zero temperature~\cite{balents} but develop topological order 
due to the long-range entanglement in the system.~\cite{wen1990} Various QSL have been suggested 
as the ground state of some frustrated magnetic systems,~\cite{balents} and have been searched 
for many years in both experimental and theoretical studies. The Kagome antiferromagnet is the 
most promising system for hosting QSL.~\cite{lee2008,mendels,han2012,marston,hastings,balents2002,wang,hermele,ymlu,lhuillier,ran,yasir,mei,jiang2008,white,schollwock,jiang,ssgong13,yche,ludwig}
In the corresponding Heisenberg model on the Kagome lattice with nearest-neighbor interactions, 
a time-reversal symmetric QSL has been discovered by different advanced numerical 
methods, with gapped~\cite{jiang2008,white,schollwock,jiang} or gapless excitations.~\cite{ran,yasir}

A sub-class of QSL, which breaks time-reversal symmetry, is called chiral spin liquid 
(CSL).~\cite{kl1989,wen1989,yang,haldane} By doping the CSL, the condensation of the anyonic 
quasi-particles might realize exotic superconductivity.~\cite{wen1989,laughlin88,wilczek}
The simplest CSL is given by the Kalmeyer-Laughlin state, which was proposed as the $\nu=1/2$ 
fractional quantum Hall state in frustrated magnetic systems.~\cite{kl1989} However, the
realization of CSL by a spontaneous time-reversal symmetry breaking in realistic frustrated 
magnetic systems was elusive in past.~\cite{thomale,cirac} Recently, the state-of-art 
density-matrix renormalization group (DMRG) has been implemented to study different spin-$1/2$ 
antiferromagnets on Kagome lattice with the spin couplings up to the third neighbors.~\cite{ssgong13,yche} 
A CSL has been suggested as the $\nu=1/2$ Laughlin state in these systems, based upon the 
calculation of the fractionally quantized Chern number $C=1/2$~\cite{ssgong13} and chiral edge 
spectrum.~\cite{yche} Meanwhile, the same CSL has been also obtained in the Heisenberg model 
with explicit time-reversal symmetry breaking chiral interactions on the Kagome lattice.~\cite{ludwig}

In theoretical studies, Wen {\it et al.}~\cite{wen1989} described the CSL states through the 
fluxes of an underlying gauge field theory within the fermionic representation, which had shed 
light on the understanding of the topological order of the CSL, including the topological 
degeneracy and fractionalized quasi-particles.~\cite{wen1990,zee1984,laughlin83,arovas,wen1991}
Recently, Zhang {\it et al.}~\cite{zhangyi} revealed the semionic statistics of quasi-particles
for the CSL state on a square lattice using the Gutzwiller projective fermionic representation
with the $\pi$-flux phase.~\cite{wen1989,ludwig94}
Motivated by the discovery of the CSL in extended Kagome systems,~\cite{ssgong13,yche} recent 
variational studies based on the Gutzwiller projected parton wave function found that the 
third-neighbor coupling could stabilize the CSL in the Heisenberg model on the Kagome 
lattice.~\cite{mei} This finding stimulates a deeper study and characterization (e.g., 
topological properties) of variational wave functions. In particular, it is interesting to 
compare the topological nature of such a CSL in the variational approach with the DMRG 
results.~\cite{ssgong13,yche,ssgong2014}

In this paper, we consider both the $J_1{-}J_2{-}J_3$ Heisenberg model:
\begin{equation}\label{ham1}
H=J_1\sum_{\langle ij\rangle} \vec{S}_{i} \cdot \vec{S}_{j} + J_2\sum_{\langle\langle ij\rangle\rangle} \vec{S}_{i} \cdot \vec{S}_{j} +J_3\sum_{\langle\langle\langle ij\rangle\rangle\rangle} \vec{S}_{i} \cdot \vec{S}_{j},
\end{equation}
and the $J_1{-}J_{\chi}$ model (with explicit chiral interactions):
\begin{equation}\label{ham2}
H_{\chi}=J_1\sum_{\langle ij\rangle} \vec{S}_{i} \cdot \vec{S}_{j} + J_{\chi}\sum_{\bigtriangleup/\bigtriangledown} \vec{S}_{i} \cdot (\vec{S}_{j}\times\vec{S}_{k}),
\end{equation}
on the Kagome lattice. In the $J_1{-}J_2{-}J_3$ model, the system has the first- ($J_1$), second- 
($J_2$) and third-neighbor ($J_3$) couplings (the latter ones, only inside each hexagon); while 
in the $J_1{-}J_{\chi}$ model, it has the chiral couplings in each up ($\bigtriangleup$) and down 
($\bigtriangledown$) triangles, and the sites $i$, $j$, and $k$ follow the clockwise order in the 
triangles. In the following, we will take $J_1=1$ as the unit of energies.

Variational wave functions are constructed by projecting mean-field states in the fermionic
representation. Through careful optimization of the variational parameters and simulations on 
large clusters, we compare the energies of the U(1) Dirac spin liquid (DSL) and CSL.
For the $J_1{-}J_2{-}J_3$ model, the CSL overcomes the DSL when $J_3$ is slightly larger 
than $J_2$. We would like to mention that, with even larger values of $J_3$, DMRG calculations 
suggested that the CSL has a transition to another spin ordered phase,~\cite{ssgong2014} but this 
issue is not addressed in the present paper. For the $J_1{-}J_{\chi}$ model, a consistent energy 
gain is obtained for the CSL state at $J_{\chi}=0.15$. The chiral-chiral correlation functions show
a long-range chiral order, consistent with DMRG results. Most importantly, the calculation of the 
topological Chern number~\cite{haldane} allows us to conclude that the CSL in both models is 
equivalent to the $\nu=1/2$ Laughlin state, as established in DMRG calculations.~\cite{ssgong13,yche}
Finally, we show that also the ground-state degeneracy, obtained by considering different boundary
conditions in the mean-field Hamiltonian, is consistent with what expected from a CSL.

\begin{figure}[b!]
\begin{center}
\includegraphics[width=\columnwidth]{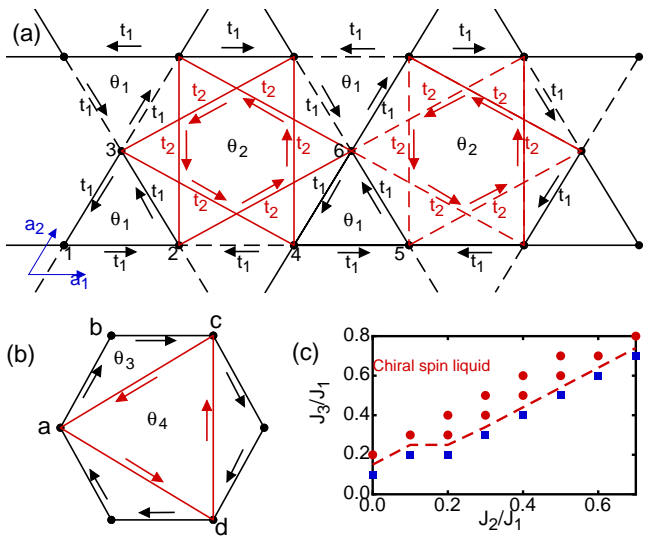}
\end{center}
\caption{ \label{0pifluxnnn}
(Color online) (a) and (b) The variational {\it Ansatz} with the NN hopping $t_{1}$ (black) and 
NNN hopping $t_{2}$ (red) is shown. Solid (dashed) lines indicate positive (negative) hoppings, 
which define the U(1) DSL. The phases $\phi_{1}$ and $\phi_{2}$ are added upon this {\it Ansatz} 
to obtain a CSL. The direction of arrows indicates one possible convention of phases. In each up 
and down triangle, the flux is $\theta_1=3\phi_1$; in each hexagon the flux is $\theta_2=\pi-6\phi_1$. 
The triangle $abc$ has flux $\theta_3=\pi-2\phi_{1}-\phi_{2}$, and the triangle $acd$ has flux 
$\theta_4=3\phi_{2}$. (c) Phase diagram: the red dots (CSL) and blue squares (U(1) DSL) are the 
calculated data. The red dashed line indicates the approximate phase boundary between the CSL and 
U(1) DSL.}
\end{figure}

{\it Method and Wave Function --}
The variational wave functions are defined by the projected mean-field states:
\begin{eqnarray}\label{meanfield}
|\Psi_{V}\rangle=\mathcal{P}_{G}|\Psi_{MF}\rangle,
\end{eqnarray}
where $\mathcal{P}_{G}=\prod_{i}(1-n_{i\uparrow}n_{i\downarrow})$ is the Gutzwiller projector, 
which enforces no double occupation on each site. $|\Psi_{MF}\rangle$ is the ground state of a 
mean-field Hamiltonian that only contains hopping:
\begin{eqnarray}\label{meanfieldham}
H_{MF}=\sum_{ij,\sigma} t_{ij}c^{\dag}_{i\sigma}c_{j\sigma} + h.c.,
\end{eqnarray}
where $c^{\dag}_{i,\sigma}$ ($c_{i,\sigma}$) creates (destroys) an electron on site $i$ with
spin $\sigma$. Different spin-liquid phases can be described by the different patterns of 
$t_{ij}$ on the bonds of the lattice. Here, we consider the hoppings for nearest-neighbor (NN) 
and next-nearest neighbor (NNN) bonds, indicated by $t_{1}$ and $t_{2}$, respectively. 
Since we are interested in CSL, we allow for both real and imaginary parts in the hopping i.e., 
$t_{ij}=|t_{ij}| e^{i\phi_{ij}}$. In Fig.~\ref{0pifluxnnn}, we show the {\it Ansatz} of our 
variational wave function; since $t^{*}_{ij} = t_{ji}$ an orientation of the bond is needed:
for the hopping from $i$ to $j$, $t_{ij}$ ($t^{*}_{ij}$) is taken in (opposite to) the direction 
of the arrow.

Here, we choose the case where the up and down triangles have the same fluxes 
(i.e., $\theta_1=3\phi_{1}$), and the flux in the hexagon is $\theta_2=\pi-6\phi_{1}$.
This state can be represented as $[3\phi_1, \pi-6\phi_1]$, as considered in Refs.~\onlinecite{ran} 
and~\onlinecite{mei}. When including also the NNN bonds, a more complex flux structure appears in 
the hexagon, as shown in Fig.~\ref{0pifluxnnn}: the triangles $abc$ have flux 
$\theta_3=\pi-2\phi_{1}-\phi_{2}$, and the triangles $acd$ have flux $\theta_4=3\phi_{2}$. 
Thus, the variational state can be represented by using the four fluxes ($\theta_{i}$) as 
$[3\phi_1, \pi-6\phi_1; \pi-2\phi_1-\phi_2, 3\phi_2]$. The U(1) DSL, which has two Dirac point (for 
each spin), has fluxes $[0,\pi,\pi,0]$; otherwise, the wave function describes a CSL.~\cite{hastings}
In our calculations, we set the real part of the NN hopping $Re(t_{1})=1$, and tune the imaginary 
part $Im(t_{1})$ to change $\phi_1$. For each $\phi_1$, we optimize the other two parameters (i.e., 
$Re(t_2)$ and $Im(t_2)$) using variational Monte Carlo to find the energetically favored state. 
In particular, we use the stochastic reconfiguration (SR) optimization method,~\cite{sorella} 
which allows us to obtain an extremely accurate determination of variational parameters. 

{\it Results --}
We performed our variational calculations for the mean-field Hamiltonian Eq.~(\ref{meanfieldham})
at half filling on toric clusters with $L \times L \times 3$ sites under the antiperiodic 
boundary conditions (APBC), and compared the U(1) DSL and the CSL. We start from the U(1) DSL, 
and add the fluxes gradually through increasing $\phi_{1}$ to get the CSL. If we only consider 
the NN hopping term $t_{1}$ within the variational wave function, we find that, for $J_2=J_3>0.3$, 
the CSL appears in the $J_1{-}J_2{-}J_3$ Heisenberg model as a local minimum. Most importantly, 
only when the NNN term $t_{2}$ is taken into account, the CSL has an energy gain with respect to 
the U(1) DSL. Therefore, in the following, we use the wave function including both $\phi_1$ and 
$\phi_2$.

\begin{figure}[b!]
\begin{center}
\includegraphics[width=\columnwidth]{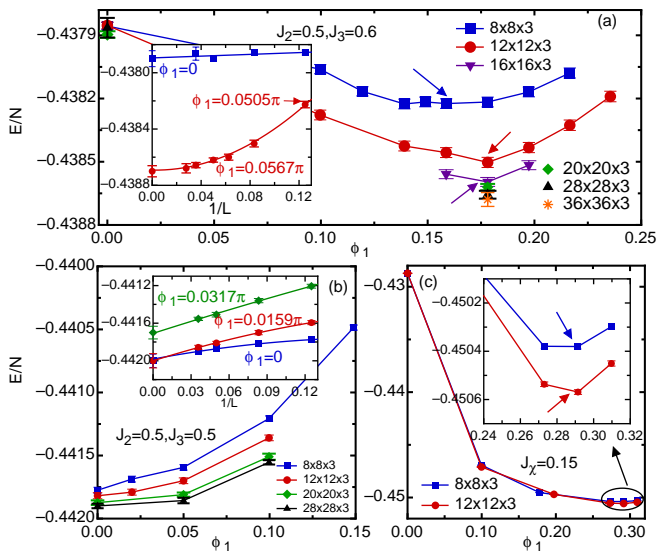}
\end{center}
\caption{ \label{diffj2j3}
(Color online) The energy per site as function of $\phi_{1}$ on different lattices. Results for the 
$J_1{-}J_2{-}J_3$ Heisenberg model at $J_2=0.5$ with $J_3=0.6$ (a) and $0.5$ (b), for the 
$J_1{-}J_{\chi}$ model at $J_{\chi}=0.15$ (c). Insets: the energy per site as function of $1/L$ for 
the U(1) DSL ($\phi_{1}=0$) up to $L=28$, and the CSL with $\phi_{1}=0.0505\pi$ on $L=8$ lattice 
and $\phi_{1}=0.0567\pi$ on larger clusters up to $L=36$ (a); the finite size effect for different 
$\phi_{1}$ (b) and with enlarged scale around $\phi_1=0.0928\pi$ (c). The arrows in (a) and (c) 
show the energy minimum, and indicate the CSL stabilized in both models.}
\end{figure}

Our main results are presented in Fig.~\ref{diffj2j3}. For the $J_1{-}J_2{-}J_3$ 
Heisenberg model of Eq.~(\ref{ham1}), we find that the CSL is energetically favored when 
$J_3$ is a little larger than $J_2$. As an example in Fig.~\ref{diffj2j3}(a), we show that,
at $J_2=0.5$ and $J_3=0.6$, the energy exhibits a minimum at a finite value of $\phi_1$, which 
is $\phi_{1}=0.0505\pi$ for $L=8$ and $\phi_{1}=0.0567\pi$ for $L=12$ and $16$. It is quite 
time consuming to perform SR optimization on larger sizes, but, fortunately, the variational 
parameters are only slightly modified from $L=12$ to $L=16$ (see supplemental material). 
Therefore, we take the wave function optimized for $L=16$ to calculate variational energies up 
to $L=36$. After the finite-size scaling, which is shown in the inset of Fig.~\ref{diffj2j3}(a),
the estimated energy per site at $J_2=0.5$ and $J_3=0.6$ is $E=-0.4387(1)$. In this case, the 
accuracy is about $5.8\%$, compared with the DMRG data on cylinder geometries (where $E=-0.465603$). 
By contrast, at $J_2=J_3=0.5$ and up to $L=12$, the best energy is given by the U(1) DSL, see 
Fig.~\ref{diffj2j3}(b). However, when we take the optimized wave functions at each $\phi_1$ and 
perform the calculations up to $L=28$, we find that the difference between the energies at 
$\phi_{1}=0$ and $0.0159\pi$ is very small (i.e., of the order of $10^{-4}$). Actually, performing
the finite size scaling yields the same estimated energy per site $E=-0.4420(1)$, as shown in the 
insert of Fig.~\ref{diffj2j3}(b). This point is very close to the boundary of the phase transition,
thus it is hard to distinguish the CSL from the U(1) DSL. More results for different values of 
$J_2$ and $J_3$ are shown in the supplemental material. 

The rough phase diagram for the $J_1{-}J_2{-}J_3$ Heisenberg model is presented in 
Fig.~\ref{0pifluxnnn}(c). Here, for $J_2=0$, we get the CSL for $J_3 \geq 0.2$, which is different 
from the conclusion in Ref.~\onlinecite{mei}, which obtained $J_3>0.3$. The reason of this 
discrepancy might be due to the energy gain obtained by including the NNN hopping $t_2$ in the 
variational wave function.

\begin{figure}[b!]
\begin{center}
\includegraphics[width=\columnwidth]{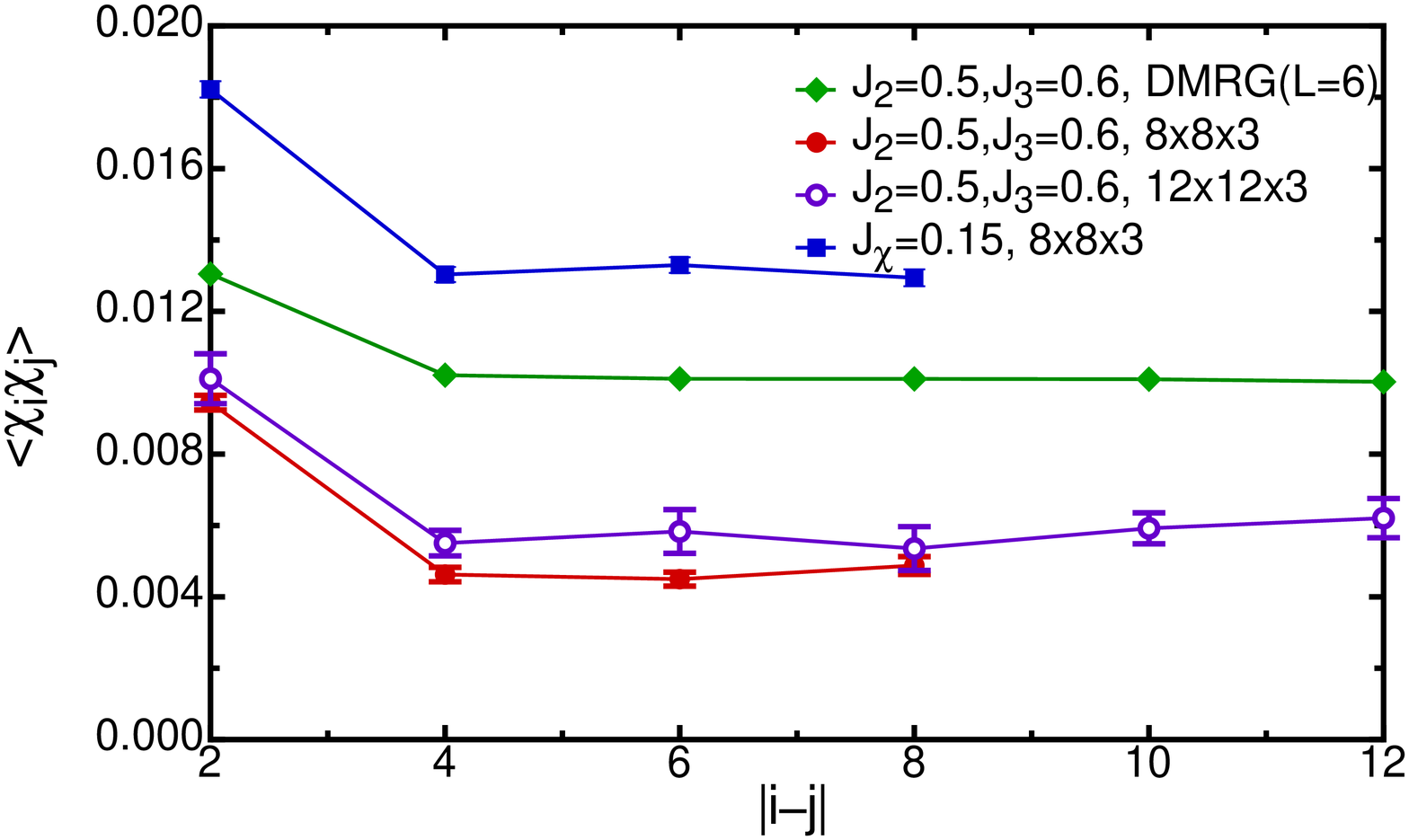}
\end{center}
\caption{ \label{chiralcorr}
(Color online) The chiral-chiral correlation function along $\vec{a}_{1}$ direction on $L=8$ 
and $12$ lattices for the $J_1{-}J_2{-}J_3$ Heisenberg model (at $J_2=0.5$ and $J_3=0.6$) and the 
$J_1{-}J_{\chi}$ model (at $J_{\chi}=0.15$). The DMRG data are calculated on cylinder with 
$L=6$ at $J_2=0.5$ and $J_3=0.6$.}
\end{figure}

In order to detect the chiral order in the optimized wave functions, we measure the chiral-chiral 
correlation function between two triangles defined as:
\begin{equation}\label{defchi}
\langle\chi_{i}\chi_{j}\rangle=\langle[\vec{S}^{i}_{1} \cdot (\vec{S}^{i}_{2}\times\vec{S}^{i}_{3})][\vec{S}^{j}_{1} \cdot (\vec{S}^{j}_{2}\times\vec{S}^{j}_{3})]\rangle,
\end{equation}
where $\chi_{i}=\vec{S}^{i}_{1} \cdot (\vec{S}^{i}_{2}\times\vec{S}^{i}_{3})$ is the chirality 
of triangle $\bigtriangleup^{i}_{123}$. In Fig.~\ref{chiralcorr}, we show the chiral-chiral 
correlation $\langle\chi_{i}\chi_{j}\rangle$ as a function of the distance $|i-j|$ between two 
up triangles at $J_2=0.5$ and $J_3=0.6$ on $8\times8\times3$ and $12\times12\times3$ lattices. 
On both clusters, $\langle\chi_{i}\chi_{j}\rangle$ decays rapidly to a finite value,
indicating the long-range chiral order, the difference between $L=8$ and $12$ being small. 
It is interesting to note that the chiral order is larger in the accurate DMRG calculations
(performed  cylinder with $L=6$) than in variational Monte Carlo ones. 

The variational state with non-trivial fluxes ($[3\phi_1, \pi-6\phi_1; \pi-2\phi_1-\phi_2, 3\phi_2]$)
can be also implemented to the $J_1{-}J_{\chi}$ model of Eq.~(\ref{ham2}). In this case, the CSL is
stabilized much easier: even for a small value of $J_{\chi}$, namely $J_{\chi}=0.15$, there is
a clear minimum in the energy around $\phi_{1}=0.0928\pi$, see Fig.~\ref{diffj2j3}(c).
For $L=12$, the CSL has an energy per site $E=-0.45057(1)$, much lower than $E=-0.42872(2)$ of the 
U(1) DSL. Also the chiral-chiral correlation function in Fig.~\ref{chiralcorr} indicates a robust 
chiral order at $J_{\chi}=0.15$.
\begin{figure}[b!]
\begin{center}
\includegraphics[width=\columnwidth]{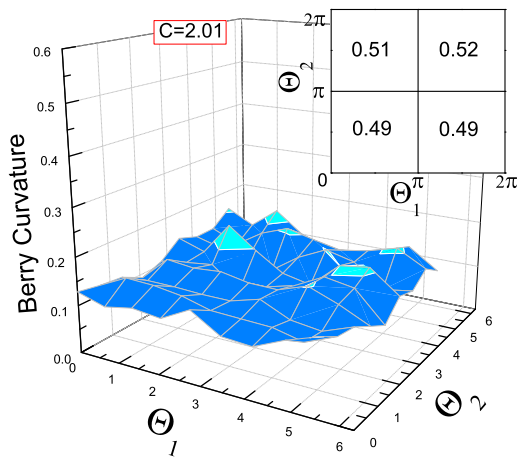}
\end{center}
\caption{ \label{berry}
(Color online) The Berry curvature at $J_2=0.5$ and $J_3=0.6$ on $L=12$ lattice. The Brillouin zone 
is divided into a mesh with 100 plaquettes. The summation between $0$ and $2\pi$ gives $C=2.01$.}
\end{figure}

Until now, we have shown that the chiral state arises in these two models in view of the 
chiral-chiral correlation. However, a CSL is further characterized by the non-trivial topological
structures, including the topological Chern number~\cite{haldane} and the degeneracy of the 
ground state.~\cite{wen1990} In the following, we proceed in these two directions to clarify 
that our variational state indeed represents a CSL state.

First, the topological Chern number is computed as the integral over the Berry curvature 
$F(\Theta_{1},\Theta_{2})$:~\cite{niu1985,sheng2003,xwan,hafezi}
\begin{equation}\label{chern}
C=\frac{1}{2\pi}\int d\Theta_{1}d\Theta_{2}F(\Theta_{1},\Theta_{2}),
\end{equation}
where $0\leq\Theta_{k}\leq2\pi$ ($k=1,2$). To compute the Chern number numerically, we consider 
twisted boundary conditions in the mean-field Hamiltonian, namely 
$c_{j+L_{k}\uparrow}=c_{j\uparrow}e^{i\Theta_{k}}$ and 
$c_{j+L_{k}\downarrow}=c_{j\downarrow}e^{-i\Theta_{k}}$. Then, we divide the 
the Brillouin zone into $M$ plaquettes, with the Berry curvature being 
$F_{l}=arg\prod_{i}\langle \Psi^{l_{i+1}}_{V}|\Psi^{l_{i}}_{V}\rangle$ 
($l=1, \dots, M$; $i=1,2,3,4$ for the four corners of the $l$-$th$ plaquette, and 
$\Psi^{l_5}_{V}=\Psi^{l_1}_{V}$), where $|\Psi^{l}_{V}\rangle$ is the projected ground state of 
the mean-field Hamiltonian with twisted boundary conditions. The overlap
$\langle \Psi^{l_{i+1}}_{V}|\Psi^{l_{i}}_{V}\rangle=\sum_{x}P(x)\frac{\langle x|\Psi^{l_{i}}_{V}\rangle}{\langle x|\Psi^{l_{i+1}}_{V}\rangle}$
is calculated by Monte Carlo method according to the weight 
$P(x)=\frac{|\langle x|\Psi^{l_{i+1}}_{V}\rangle|^{2}}{\sum_{x}|\langle x|\Psi^{l_{i+1}}_{V}\rangle|^{2}}$.
In order to numerically check the accuracy, we changed $M$ from 100 up to 400 plaquettes. 
The numerical results show that the dimension of mesh changes the Berry curvature, but gives the same 
topological Chern number. We must emphasize that the integration from $0$ to $2\pi$ in the twist of 
the fermionic operators includes two periods of phases for the spin operators and, therefore,
the result must be divided by $4$. The integration between $0$ and $2\pi$ gives $2$ with high 
accuracy (see Fig.~\ref{berry}), leading to $C=1/2$, which is fully consistent with recent DMRG 
results.~\cite{ssgong13}

The degeneracy of the wave function, which indicates non-trivial ground-state structure, was not 
obvious to obtain from the variational approach with the Gutzwiller projected parton 
construction. Recently, Zhang {\it et al.}~\cite{zhangyi} realized that the ground-state 
degeneracy is consistent with the linear dependence for variational wave functions through 
fermionic construction for SU(2) Chern-Simons theory. We follow the same idea to construct the 
linearly independent states from the four projected states that are obtained through changing 
the boundary conditions of the mean field Hamiltonian in $\vec{a}_{1}$ and $\vec{a}_{2}$ 
directions (see Fig.~\ref{0pifluxnnn}). We denote the four states as $|\psi_{1},\psi_{2}\rangle$,
where $\psi_{i}=0$ for periodic boundary condition and $\pi$ for antiperiodic boundary condition
($i=1,2$). In order to find the linearly independent states from these four projected states 
(i.e., $\{|0,0\rangle, |0,\pi\rangle, |\pi,0\rangle, |\pi,\pi\rangle\}$), we calculate the overlaps 
between all possible states. The numerical calculations up to $L=32$ indicate strong size effects, 
due to the smallness of the mean-field gap. For example, for $J_2=0.5$ and $J_3=0.6$ on the 
$16\times16\times3$ lattice with $\phi_1=0.0567\pi$ and $\phi_2=-0.2807\pi$, the band gap is only 
about $0.6$.

Thus, in order to suppress the finite-size effects on small clusters, we tune $\phi_1$ and $\phi_2$ 
to enlarge the mean-field band gap (up to a value of $2$), still keeping the long-range chiral 
order and a non-zero Chern number. Consequently, even on the small $8\times8\times3$ lattice, we 
find the two linear independent states indicating the two-fold degeneracy,
which are (see the supplemental material for details):
\begin{eqnarray}
&&|1\rangle=|00\rangle=\frac{1}{\sqrt{3}}(|0,\pi\rangle+|\pi,0\rangle+|\pi,\pi\rangle), \\
&&|2\rangle=\frac{1}{\sqrt{6}}(|0,\pi\rangle+e^{i\frac{2\pi}{3}}|\pi,0\rangle+e^{-i\frac{2\pi}{3}}|\pi,\pi\rangle).
\end{eqnarray}

{\it Conclusions --}
In conclusion, we investigated the CSL in the $J_1{-}J_2{-}J_3$ Heisenberg models on the Kagome
lattice by the variational approach with Gutzwiller projected fermionic construction. Our 
variational studies reveal that the CSL is energetically favored in the phase region consistent 
with the recent DMRG calculations.~\cite{ssgong13} However, differently from the DMRG results, 
we found that when $J_3$ is larger than $J_2$, instead of $J_2=J_3$, the CSL begins to appear 
as indicated by the existence of energy minimum while tuning the fluxes. 
On the other hand, we have shown that our wave function also works for the $J_1-J_{\chi}$ model,
which includes a three-spin parity and time reversal violating interaction.

Further investigations with these optimized wave functions show that the spin-spin correlation 
functions decay fast (see supplemental material), indicating short-range correlations. Instead,
the chiral-chiral correlation function presents a long-range chiral order, consistent with DMRG 
results. The variational wave function underestimates the robust of the CSL, as the accurate 
DMRG calculations show stronger chiral order than variational state. Moreover, our calculations 
of the topological Chern number and the ground state degeneracy suggest that the chiral state is 
the $\nu=1/2$ Laughlin state.

{\it Acknowledgements --}
W.-J.H. and F.B. thank D. Poilblanc and Y. Iqbal for useful discussions.
This research is supported by the National Science Foundation through grants DMR-1408560 (W.-J.H, 
D.N.S),  DMR-1205734 (S.S.G.), and U.S. Department of Energy, Office of Basic Energy Sciences 
under grant No. DE-FG02-06ER46305 (W.Z.). Y.Z. is supported by SITP. F.B. by PRIN 2010-11.


\clearpage
\begin{center}
\begin{large}
{\bf Supplemental Material}
\end{large}
\end{center}

\begin{figure*}[b!]
\begin{center}
\includegraphics[width=1.5\columnwidth]{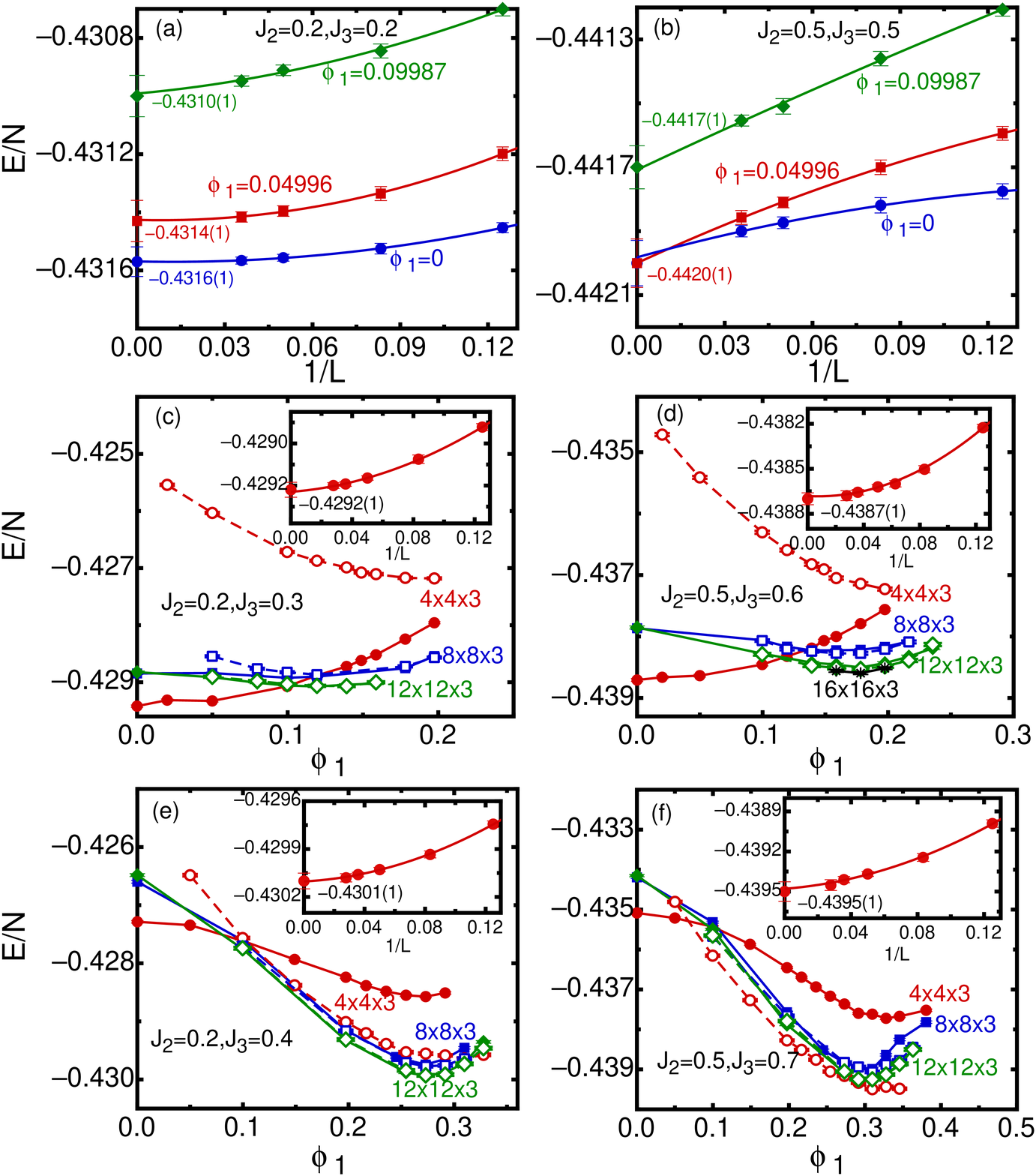}
\end{center}
\caption{ \label{diffj2j3bc}
(Color online) The energy per site on different lattices. The finite-size scaling and different flux $\phi_1$ is shown for 
$J_3=J_2=0.2$ (a) and $J_3=J_2=0.5$ (b). Energy per site as a function of $\phi_1$ for different values of $J_2$ and
$J_3$ are reported in (c), (d), (e), and (f), for various sizes of the cluster: $4\times4\times3$ (red circles),
$8\times8\times3$ (blue squares), $12\times12\times3$ (green diamonds), and $16\times16\times3$ (black stars). 
The filled (empty) points with solid (dashed) lines indicate APBC (PBC). The insets show the finite size scaling.}
\end{figure*}

{\it More energy data and spin-spin correlation functions ---} 
We show the energy per site at different values of $J_2$ and $J_3$ with antiperiodic (APBC) and periodic (PBC) 
boundary conditions in Fig.~\ref{diffj2j3bc}. On the $4\times4\times3$ lattice, the results depend on different 
boundary conditions, owing to finite-size effect. As the size is increased to the $8\times8\times3$ lattice, the 
energy per site only exhibits slight difference under different boundary conditions (of the order of $10^{-4}$). 
On larger clusters, the energy results with different boundary conditions are the same within one error bar. 
The optimized values of the phases $\phi_1$ and $\phi_2$ for the wave functions with APBC and PBC are reported 
in Table~\ref{wfflux1}. By taking the wave functions optimized on the $12\times12\times3$ or $16\times16\times3$ 
lattice, we perform variational calculations up to $36\times36\times3$ lattice, and perform the finite-size scaling 
to obtain the estimated energies (see the insets of Fig.~\ref{diffj2j3bc}).

\begin{figure*}[b!]
\begin{center}
\includegraphics[width=1.5\columnwidth]{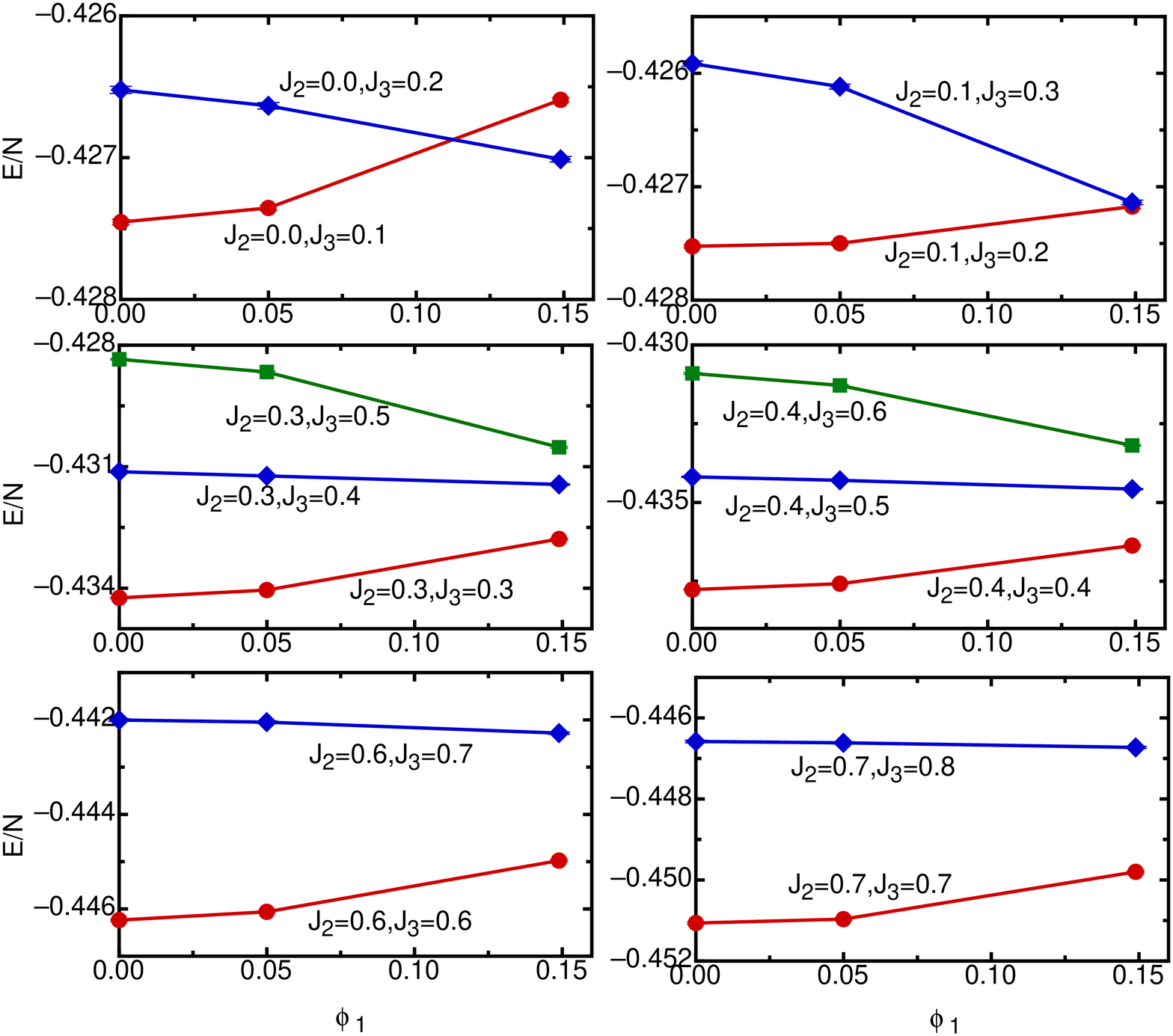}
\end{center}
\caption{ \label{diffj2j3L8}
(Color online) The energy per site at different $J_{2}$ and $J_{3}$ on the $8\times8\times3$ lattice. A CSL is energetically favored
when the energy per site shows a minimum as a function of $\phi_1$.}
\end{figure*}

Fig.~\ref{diffj2j3L8} shows several calculations on the $8\times8\times3$ lattice with APBC at different $J_2$ 
and $J_3$, qualitatively similar results are obtained with different boundary conditions. The CSL is energetically
favored when the energy per site decreases as the flux $\phi_1$ is increased. The schematic phase diagram shown
in the main text is constructed from these results.

\begin{figure*}[b!]
\begin{center}
\includegraphics[width=1.2\columnwidth]{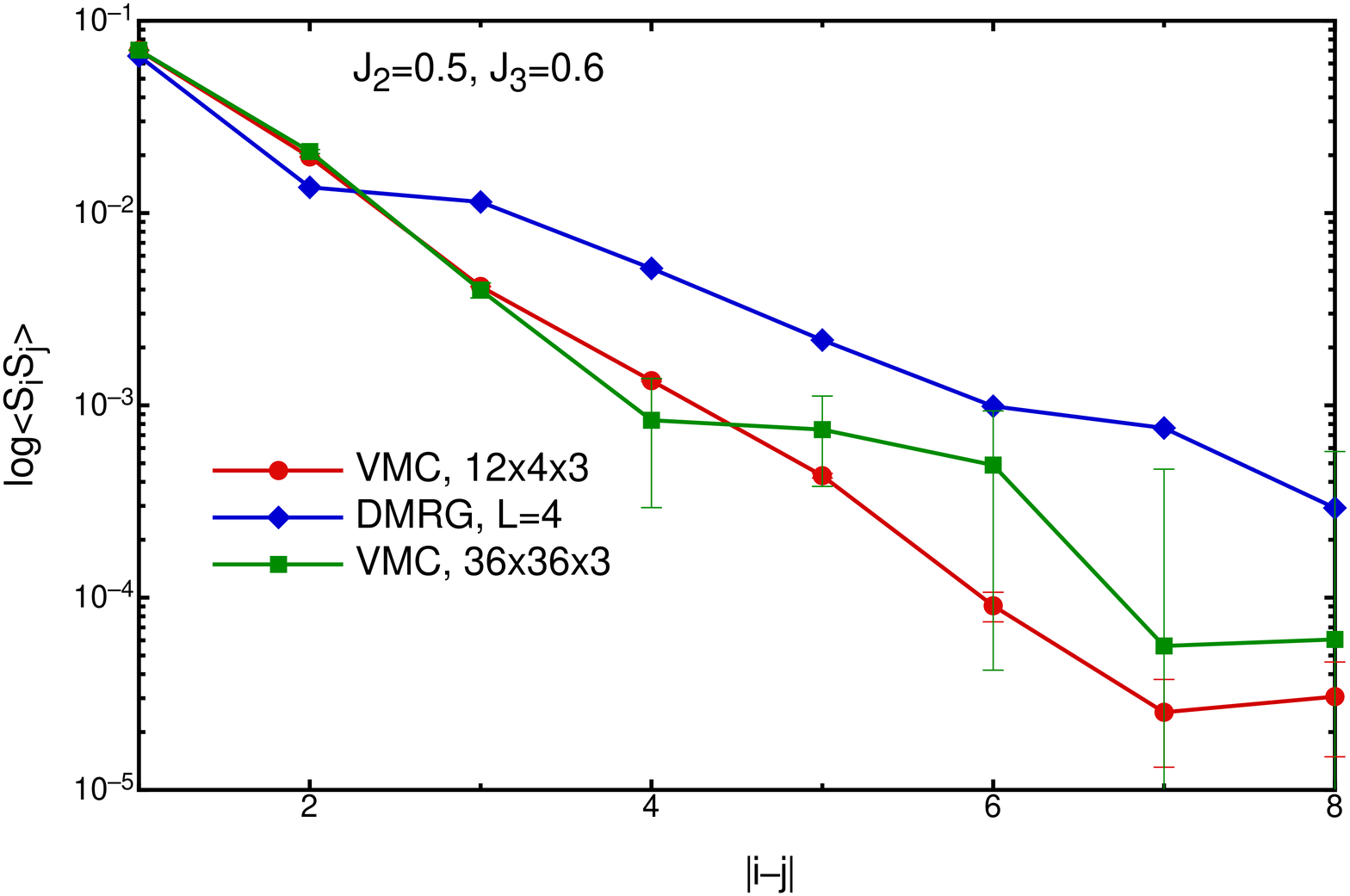}
\end{center}
\caption{ \label{spin}
(Color online) The spin-spin correlation function along the $\vec{a}_1$ direction for $J_{2}=0.5$ and $J_3=0.6$ is shown for the
variational Monte Carlo (VMC) and density-matrix renormalization group (DMRG) calculations.}
\end{figure*}

Taking the optimized variational wave function on the $16\times16\times3$ lattice, we measure the spin-spin 
correlation function, see Fig.~\ref{spin} that also shows a comparison with DMRG results on cylinder with $L=4$. 
The fast decay of the spin correlation indicates the absence of a long-range order, consistent with DMRG conclusions.

{\it Details for the degeneracy ---} 
In order to find the linearly independent ground states, we numerically calculate all the overlaps between states
with different boundary conditions in the mean-field Hamiltonian, which are indicated by
$\{|\psi_{1},\psi_{2}\rangle\}=\{|0,0\rangle, |0,\pi\rangle, |\pi,0\rangle, |\pi,\pi\rangle\}$. 
In order to maximize the mean-field gap, i.e., $\Delta \simeq 2$, we take $\phi_{1}=0.1015\pi$ and $\phi_{2}=0.25\pi$
and perform the calculations on the $8\times8\times3$ lattice. By fixing the global phases in such a way
that all the overlaps with $|0,0\rangle$ are real, we get the overlap matrix ${\cal O}$:~\cite{zhangyi} 
\begin{eqnarray}\label{biggap}
{\cal O} &=&\left(\begin{array}{cccc}
\langle 0,0|0,0\rangle & \langle 0,0|0,\pi\rangle & \langle 0,0|\pi,0\rangle & \langle 0,0|\pi,\pi\rangle \\ 
\langle 0,\pi|0,0\rangle & \langle 0,\pi|0,\pi\rangle & \langle 0,\pi|\pi,0\rangle & \langle 0,\pi|\pi,\pi\rangle \\
\langle \pi,0|0,0\rangle & \langle \pi,0|0,\pi\rangle & \langle \pi,0|\pi,0\rangle & \langle \pi,0|\pi,\pi\rangle \\
\langle \pi,\pi|0,0\rangle & \langle \pi,\pi|0,\pi\rangle & \langle \pi,\pi|\pi,0\rangle & \langle \pi,\pi|\pi,\pi\rangle
\end{array}\right) \nonumber\\
&\approx&\left(\begin{array}{cccc}
1 & 0.57 & 0.57 & 0.57 \\ 
0.57 &1 & 0.57e^{-i1.6} & 0.58e^{i1.6} \\
0.57 & 0.57e^{i1.6} &1 & 0.57e^{-i1.6} \\
0.57 & 0.57e^{-i1.6} & 0.57e^{i1.5} & 1
\end{array}\right) \nonumber\\
&\approx&\left(\begin{array}{cccc}
1                  & \frac{1}{\sqrt{3}}  & \frac{1}{\sqrt{3}}  & \frac{1}{\sqrt{3}}  \\ 
\frac{1}{\sqrt{3}} & 1                   & -\frac{i}{\sqrt{3}} & \frac{i}{\sqrt{3}}  \\
\frac{1}{\sqrt{3}} & \frac{i}{\sqrt{3}}  & 1                   & -\frac{i}{\sqrt{3}} \\
\frac{1}{\sqrt{3}} & -\frac{i}{\sqrt{3}} & \frac{i}{\sqrt{3}}  & 1
\end{array}\right).
\end{eqnarray}
The independent ground states can be found by diagonalizing the overlap matrix, i.e., ${\cal O}=U^{\dag} \Lambda U$. 
We find that only two eigenvalues are non-zero, indicating that only two eigenvectors are linearly independent. 
This fact implies that the ground-state degeneracy is two-fold. In particular, these two states can be constructed 
such to preserve lattice symmetries. For example, we can consider the $2\pi/3$ rotations, generated by the operator 
$R_{2\pi/3}$. Within the Gutzwiller projected fermionic representation, the four wave functions from different boundary
conditions have the following relations under $R_{2\pi/3}$ rotations:
\begin{eqnarray}
R_{2\pi/3}|00\rangle&=&|00\rangle, \nonumber \\
R_{2\pi/3}|0\pi\rangle&=&|\pi0\rangle, \nonumber \\
R_{2\pi/3}|\pi0\rangle&=&|\pi\pi\rangle, \nonumber \\
R_{2\pi/3}|\pi\pi\rangle&=&|0\pi\rangle. \nonumber
\end{eqnarray}
Eigenstates of $R_{2\pi/3}$ can be easily constructed, with eigenvalues $r=1$, $r=e^{-\frac{2\pi i}{3}}$, and 
$r=e^{\frac{2\pi i}{3}}$:
\begin{eqnarray}
R_{2\pi/3}&:& r=1 \nonumber\\
&&|1\rangle=|00\rangle, \nonumber\\
R_{2\pi/3}&:& r=e^{-\frac{2\pi i}{3}}\nonumber\\
&&|2\rangle\propto|0\pi\rangle+e^{\frac{2\pi i}{3}}|\pi0\rangle+e^{-\frac{2\pi i}{3}}|\pi\pi\rangle, \nonumber\\
R_{2\pi/3}&:& r=e^{\frac{2\pi i}{3}}\nonumber\\
&&|3\rangle\propto|0\pi\rangle+e^{-\frac{2\pi i}{3}}|\pi0\rangle+e^{\frac{2\pi i}{3}}|\pi\pi\rangle, \nonumber\\
R_{2\pi/3}&:& r=1 \nonumber\\
&&|4\rangle\propto|0\pi\rangle+|\pi0\rangle+|\pi\pi\rangle. \nonumber
\end{eqnarray}

Our numerical results of Eq.~(\ref{biggap}) imply that only two linearly independent states exist:
\begin{eqnarray}
&&|1\rangle=|00\rangle, 
\label{basis1} \\
&&|2\rangle=\frac{1}{\sqrt{6}}(|0,\pi\rangle+e^{\frac{2\pi i}{3}}|\pi,0\rangle+e^{-\frac{2\pi i}{3}}|\pi,\pi\rangle),
\label{basis2} \\
&&|3\rangle=\varnothing,
\label{basis3} \\
&&|4\rangle=\frac{1}{\sqrt{3}}(|0,\pi\rangle+|\pi,0\rangle+|\pi,\pi\rangle) = |1\rangle.
\label{basis4}
\end{eqnarray}

Therefore, the four projected states $\{|\psi_{1},\psi_{2}\rangle\}$ can be represented by using the two independent
states $|1\rangle$ and $|2\rangle$: 
\begin{eqnarray}
\left|00\right\rangle     &=&\left|1\right\rangle, \nonumber\\
\left|0\pi\right\rangle   &=&\frac{1}{\sqrt{3}} \left|1\right\rangle + \sqrt{\frac{2}{3}} \left|2\right\rangle,\nonumber\\
\left|\pi0\right\rangle   &=&\frac{1}{\sqrt{3}} \left|1\right\rangle + \sqrt{\frac{2}{3}} e^{-\frac{2\pi i}{3}}\left|2\right\rangle,\nonumber\\
\left|\pi\pi\right\rangle &=&\frac{1}{\sqrt{3}} \left|1\right\rangle + \sqrt{\frac{2}{3}} e^{\frac{2\pi i}{3}} \left|2\right\rangle. \nonumber
\end{eqnarray}

{\it Relation between the ground states and the minimum entropy states ---}
In this part, we want to find the relation between the two linearly independent states $|1\rangle$ and $|2\rangle$ of
Eqs.~(\ref{basis1}) and~(\ref{basis2}) and the minimum entropy states (MESs).~\cite{zhangyi} The MESs are the useful 
basis of the degenerate ground-state manifold for topological ordered phases and label the eigenstates with different 
quasiparticles threaded through the non-contractible loop along a given direction. The transformation between the 
MES bases along different directions connected by a symmetry rotation is encoded in the corresponding modular 
matrix.~\cite{zhangyi}

Since the Kagome lattice is symmetric under $2\pi/3$ rotation $R_{2\pi/3}$ and the topological order involve no 
symmetry breaking, the overall ground-state manifold is invariant under $R_{2\pi/3}$. Nevertheless, each individual ground
state may still transform differently under $R_{2\pi/3}$, which can be interpreted as a conformal transformation:
$\left(\vec{w}_{1},\vec{w}_{2}\right)^{T}\rightarrow\left(\vec{w}_{2},\vec{w}_{3}\right)^{T}=\left(\vec{w}_{2},-\vec{w}_{1}-\vec{w}_{2}\right)^{T}$,
where
\begin{eqnarray}
\vec{w}_{1}&=&\hat{x}, \nonumber \\
\vec{w}_{2}&=&\left(-\hat{x}+\sqrt{3}\hat{y}\right)/2, \nonumber \\
\vec{w}_{3}&=&\left(-\hat{x}-\sqrt{3}\hat{y}\right)/2, \nonumber
\end{eqnarray}
are three directional vectors along the Bravais lattice vectors. 

Therefore, on the Kagome lattice, the $R_{2\pi/3}$ rotation leads to the modular transformation $\mathcal{SU}$ on 
MESs.~\cite{zhangyi} For the CSL phase (e.g., the $\nu=1/2$ Laughlin state), we have the modular $\mathcal{SU}$ matrix:
\begin{eqnarray}
\mathcal{S}=\frac{1}{\sqrt{2}}\left(\begin{array}{cc}
1 & 1\\
1 & -1
\end{array}\right), \\
\mathcal{U}=\left(\begin{array}{cc}
1 & 0\\
0 & i
\end{array}\right)e^{-\frac{2\pi i}{24}},
\end{eqnarray}
where the phase factor in the definition of $\mathcal{U}$ is given by the chiral edge central charge. The first and 
the second columns (rows) are the identity particle and the semion quasi-particle, respectively. Physically, for Abelian 
topological orders, the modular $\mathcal{S}$ matrix determines the mutual statistics of a given quasi-particle encircling 
around another one, while the modular $\mathcal{U}$ matrix contains the self-statistics of the each quasi-particle.

With the modular $\mathcal{SU}$ matrix as the $2\pi/3$ rotation operator, we can obtain the MESs along the 
$\vec{w}_{2}$ and $\vec{w}_{3}$ directions $\Xi_{1,\vec{w}_2}$ and $\Xi_{1,\vec{w}_3}$ in terms of the MESs along 
the $\vec{w}_{1}$ direction $\Xi_{1,\vec{w}_1}$, i.e., $\Xi_{1,\vec{w}_2}={\mathcal SU}\Xi_{1,\vec{w}_1}$ and
$\Xi_{1,\vec{w}_3}={\mathcal SU}\Xi_{1,\vec{w}_2}$. Similarly, we can construct the other set of MESs $\Xi_{s,\vec{w}_1}$,
$\Xi_{s,\vec{w}_2}$, and $\Xi_{s,\vec{w}_3}$:
\begin{center}
\begin{tabular}{|c|c|c|}
\hline
MES direction        & $\Xi_{1,\vec{w}_i}$                           & $\Xi_{s,\vec{w}_i}$                           \tabularnewline
\hline $\vec{w}_{1}$ & $\left(1,0\right)^{T}$                        & $\left(0,1\right)^{T}$                        \tabularnewline 
\hline $\vec{w}_{2}$ & $\left(1,1\right)^{T}e^{-\pi i/12}/\sqrt{2}$  & $\left(1,-1\right)^{T}e^{5\pi i/12}/\sqrt{2}$ \tabularnewline 
\hline $\vec{w}_{3}$ & $\left(i,1\right)^{T}e^{-5\pi i/12}/\sqrt{2}$ & $\left(1,i\right)^{T}e^{-\pi i/12}/\sqrt{2}$  \tabularnewline \hline
\end{tabular}
\end{center}

Since under $R_{2\pi/3}$ rotations, $\vec{w}_{1}\rightarrow\vec{w}_{2}$, $\vec{w}_{2}\rightarrow\vec{w}_{3}$ and 
$\vec{w}_{3}\rightarrow\vec{w}_{1}$, we may construct its eigenstates as the following:
\begin{eqnarray}
R_{2\pi/3}&:& r=1: \nonumber\\
|1\rangle_{R} &\propto&\Xi_{1,\vec{w}_{1}}+\Xi_{1,\vec{w}_{2}}+\Xi_{1,\vec{w}_{3}} \nonumber\\
&\propto&(1+\frac{e^{-\frac{\pi i}{12}}+e^{\frac{\pi i}{12}}}{\sqrt{2}},\frac{e^{-\frac{\pi i}{12}}+e^{-\frac{5\pi i}{12}}}{\sqrt{2}})^{T}\nonumber\\
&\propto&(1+\sqrt{2}\cos \frac{\pi}{12},(1-i)\cos \frac{\pi}{6})^{T}, \label{mes1} \\
R_{2\pi/3}&:& r=e^{-\frac{2\pi i}{3}}: \nonumber \\
|2\rangle_{R}
&\propto&\Xi_{1,\vec{w}_{1}}+e^{\frac{2\pi i}{3}}\Xi_{1,\vec{w}_{2}}+e^{-\frac{2\pi i}{3}}\Xi_{1,\vec{w}_{3}} \nonumber\\
&\propto&(1+\frac{e^{\frac{7\pi i}{12}}+e^{-\frac{7\pi i}{12}}}{\sqrt{2}},\frac{e^{\frac{7\pi i}{12}}+e^{\frac{11\pi i}{12}}}{\sqrt{2}})^{T} \nonumber\\
&\propto&(1-\sqrt{2} \sin \frac{\pi}{12}),(-1+i)\cos \frac{\pi}{6})^{T}, \label{mes2} \\
R_{2\pi/3}&:& r=e^{\frac{2\pi i}{3}}: \nonumber \\
|3\rangle_{R}
&\propto&\Xi_{1,\vec{w}_{1}}+e^{-\frac{2\pi i}{3}}\Xi_{1,\vec{w}_{2}}+e^{\frac{2\pi i}{3}}\Xi_{1,\vec{w}_{3}} \nonumber\\
&\propto&(1+\frac{e^{-\frac{3\pi i}{4}}+e^{\frac{3\pi i}{4}}}{\sqrt{2}},\frac{e^{-\frac{3\pi i}{4}}+e^{\frac{\pi i}{4}}}{\sqrt{2}})^{T} \nonumber \\
&=&\varnothing, \label{mes3}
\end{eqnarray}
which are consistent with the requirement that the ground states are only two-fold degenerate (similar results are 
obtained by using $\Xi_{s,\vec{w}_i}$).

From Eqs.~(\ref{basis1}), (\ref{basis2}), (\ref{mes1}), and~(\ref{mes2}), we have that:
\begin{eqnarray}
|1\rangle_{R}&=&|1\rangle \nonumber\\
&\propto&(1+\sqrt{2} \cos \frac{\pi}{12},(1-i)\cos \frac{\pi}{6})^{T} \nonumber \\
&\propto&|0,\pi\rangle+|\pi,0\rangle+|\pi,\pi\rangle\propto|0,0\rangle, \nonumber\\
|2\rangle_{R}&=&|2\rangle \nonumber\\
&\propto&(1-\sqrt{2} \sin \frac{\pi}{12},(-1+i)\cos \frac{\pi}{6})^{T} \nonumber \\
&\propto&|0,\pi\rangle+e^{\frac{2\pi i}{3}}|\pi,0\rangle+e^{-\frac{2\pi i}{3}}|\pi,\pi\rangle), \nonumber\\
|3\rangle_{R}&=&|3\rangle \nonumber\\
&\propto&|0,\pi\rangle+e^{-\frac{2\pi i}{3}}|\pi,0\rangle+e^{\frac{2\pi i}{3}}|\pi,\pi\rangle \nonumber\\
&\propto&\varnothing. \nonumber
\end{eqnarray}

\begin{table*}[!hbp]
\centering
\caption{\label{wfflux1}
We list the fluxes $\phi_{1}$ and $\phi_{2}$ from the best chiral states in Fig.~\ref{diffj2j3bc} for different 
values of $J_{2}$ and $J_{3}$ on different sizes with APBC and PBC.}
\begin{tabular}{cc|cc|cc|cc|cc}
\hline\hline
 \multicolumn{2}{c|}{$APBC$} & \multicolumn{2}{c|}{$N=48$} & \multicolumn{2}{c|}{$N=192$} & \multicolumn{2}{c|}{$N=432$} & \multicolumn{2}{c}{$N=768$} \\
\hline
$J_{1}$ & $J_{2}$ & $\phi_{1}$  & $\phi_{2}$         & $\phi_{1}$ & $\phi_{2}$            & $\phi_{1}$ & $\phi_{2}$           & $\phi_{1}$ & $\phi_{2}$           \\
\hline
  $0.2$  &   $0.3$   & $0$               & $0$                   & $0.0032\pi$ & $-0.3387\pi$          & $0.0380\pi$ & $-0.3801\pi$        &  &         \\
  $0.2$  &   $0.4$   & $0.0869\pi$ & $-0.6100\pi$   & $0.0869\pi$ & $-0.6117\pi$          & $0.0869\pi$ & $-0.6150\pi$        &  &         \\
  $0.5$  &   $0.6$   & $0$               & $0$                   & $0.0505\pi$ & $-0.2332\pi$          & $0.0567\pi$ & $-0.2802\pi$        & $0.0567\pi$ & $-0.2807\pi$        \\
  $0.5$  &   $0.7$   & $0.1043\pi$ & $-0.6071\pi$   & $0.0986\pi$ & $-0.5755\pi$          & $0.0986\pi$ & $-0.5803\pi$        &  &         \\
\hline\hline
 \multicolumn{2}{c|}{$PBC$} & \multicolumn{2}{c|}{$N=48$} & \multicolumn{2}{c|}{$N=192$} & \multicolumn{2}{c|}{$N=432$} & \multicolumn{2}{c}{$N=768$} \\
\hline
$J_{1}$ & $J_{2}$ & $\phi_{1}$  & $\phi_{2}$         & $\phi_{1}$ & $\phi_{2}$            & $\phi_{1}$ & $\phi_{2}$           & $\phi_{1}$ & $\phi_{2}$           \\
\hline
  $0.2$  &   $0.3$   & $$ & $$ & $0.0032\pi$ & $-0.3347\pi$          & $0.0380\pi$ & $-0.3805\pi$        &  &         \\
  $0.2$  &   $0.4$   & $$ & $$ & $0.0869\pi$ & $-0.6138\pi$          & $0.0869\pi$ & $-0.6154\pi$        &  &         \\
  $0.5$  &   $0.6$   & $$ & $$ & $0.0505\pi$ & $-0.2390\pi$          & $0.0567\pi$ & $-0.2785\pi$        &  &         \\
  $0.5$  &   $0.7$   & $$ & $$ & $0.0986\pi$ & $-0.5780\pi$          & $0.0986\pi$ & $-0.5805\pi$        &  &         \\
\hline\hline
\end{tabular}
\end{table*}

\begin{thebibliography}{99} 
\bibitem{balents} L. Balents, Nature (London) {\bf 464}, 199 (2010).
\bibitem{wen1990} X.-G. Wen, Phys. Rev. B {\bf 40}, 7387 (1989); X.-G. Wen, Int. J. Mod. Phys. B {\bf 4}, 239 (1990); X.-G. Wen and Q. Niu, Phys. Rev. B {\bf 41}, 9377 (1990). 
\bibitem{lee2008} P.A. Lee, Science {\bf 321}, 1306 (2008).
\bibitem{mendels} P. Mendels and F. Bert, J. Phys. Conf. Ser. {\bf 320}, 012004 (2011).
\bibitem{han2012} T.-H. Han, J. S. Helton, S. Chu, D. G. Nocera, J. A. Rodriguez-Rivera, C. Broholm, and Y. S. Lee, Nature {\bf 492}, 406 (2012).
\bibitem{marston} J. Marston and C. Zeng, J. Appl. Phys. {\bf 69}, 5962 (1991).
\bibitem{hastings} M.B. Hastings, Phys. Rev. B {\bf 63}, 014413 (2000).
\bibitem{balents2002} L. Balents, M.P.A. Fisher, and S.M. Girvin, Phys. Rev. B {\bf 65}, 224412 (2002).
\bibitem{wang} F. Wang and A. Vishwanath, Phys. Rev. B {\bf 74}, 174423 (2006).
\bibitem{hermele} M. Hermele, Y. Ran, P.A. Lee, and X.-G. Wen, Phys. Rev. B {\bf 77}, 224413 (2008).
\bibitem{ymlu} Y.-M. Lu, Y. Ran, and P.A. Lee, Phys. Rev. B {\bf 83}, 224413 (2011); Y.-M. Lu, G. Y. Cho, and A. Vishwanath, arXiv: 1403.0575.
\bibitem{lhuillier} L. Messio, B. Bernu, and C. Lhuillier, Phys. Rev. Lett. {\bf 108}, 207204 (2012).
\bibitem{ran} Y. Ran, M. Hermele, P.A. Lee, and X.-G. Wen, Phys. Rev. Lett. {\bf 98}, 117205 (2007).
\bibitem{yasir} Y. Iqbal, F. Becca, and D. Poilblanc, Phys. Rev. B {\bf 84}, 020407 (2011); Y. Iqbal, F. Becca, S. Sorella, and D. Poilblanc, Phys. Rev. B {\bf 87}, 060405 (2013); Y. Iqbal, D. Poilblanc, and F. Becca, Phys. Rev. B {\bf 89}, 020407 (2014).
\bibitem{mei} J.-W. Mei and X.-G. Wen, arXiv:1407.0869.
\bibitem{jiang2008} H.C. Jiang, Z.Y. Weng, and D.N. Sheng, Phys. Rev. Lett. {\bf 101}, 117203 (2008).
\bibitem{white} S. Yan, D. Huse, and S. White, Science {\bf 332}, 1173 (2011).
\bibitem{schollwock} S. Depenbrock, I.P. McCulloch, and U. Schollw$\ddot{o}$ck, Phys. Rev. Lett. {\bf 109}, 067201 (2012).
\bibitem{jiang} H.C. Jiang, Z. Wang, and L. Balents, Nat. Phys. {\bf 8}, 902 (2012).
\bibitem{ssgong13} S.S. Gong,  W. Zhu, and D.N. Sheng, Scientific Reports {\bf 4}, 6317 (2014).
\bibitem{yche} Y.C. He, D.N. Sheng, and Y. Chen, Phys. Rev. Lett. {\bf 112}, 137202 (2014).
\bibitem{ludwig} B. Bauer, B.P. Keller, M. Dolfi, S. Trebst, and A.W.W. Ludwig, arXiv:1303.6963; B. Bauer, L. Cincio, B.P. Keller, M. Dolfi, G. Vidal, S. Trebst, and A.W.W. Ludwig, arXiv:1401.3017.
\bibitem{kl1989} V. Kalmeyer and R.B. Laughlin, Phys. Rev. B {\bf 39}, 11879 (1989).
\bibitem{wen1989} X.-G. Wen, F. Wilczek, and A. Zee, Phys. Rev. B {\bf 39}, 11413 (1989).
\bibitem{yang} K. Yang, L.K. Warman, and S.M. Girvin, Phys. Rev. Lett. {\bf 70}, 2641 (1993).
\bibitem{haldane} F.D.M. Haldane and D.P. Arovas, Phys. Rev. B {\bf 52}, 4223 (1995).
\bibitem{laughlin88} R.B. Laughlin, Phys. Rev. Lett. {\bf 60}, 2677 (1988).
\bibitem{wilczek} F. Wilczek, Fractional Statistics and Anyon Superconductivity (World Science, Singapore, 1990).
\bibitem{thomale} D.F. Schroeter, E. Kaplt, R. Thomale, and M. Grelter, Phys. Rev. Lett. {\bf 99}, 097202 (2007).
\bibitem{cirac} A.E.B. Nielsen, J.I. Cirac, and G. Sierra, Phys. Rev. Lett. {\bf 108}, 257206 (2012).
\bibitem{zee1984} F. Wilczek and A. Zee, Phys. Rev. Lett. {\bf 52}, 2111 (1984).
\bibitem{laughlin83} R.B. Laughlin, Phys. Rev. Lett. {\bf 50}, 1395 (1983).
\bibitem{arovas} D. Arovas, J.R. Schrieffer, and F. Wilczek, Phys. Rev. Lett. {\bf 53}, 722 (1984).
\bibitem{wen1991} X.-G. Wen, Phys. Rev. B {\bf 44}, 2664 (1991).
\bibitem{zhangyi} Y. Zhang, T. Grover, and A. Vishwanath, Phys. Rev. B {\bf 84}, 075128 (2011); Y. Zhang, T. Grover, A. Turner, M. Oshikawa, and A. Vishwanath, Phys. Rev. B {\bf 85}, 235151 (2012); Y. Zhang and A. Vishwanath, Phys. Rev. B {\bf 87}, 161113 (2013). 
\bibitem{ludwig94} A.W.W. Ludwig, M.P.A. Fisher, R. Shankar, and G. Grinstein, Phys. Rev. B {\bf 50}, 7526 (1994).
\bibitem{ssgong2014} S.S. Gong, W. Zhu, L. Balents, and D.N. Sheng, in preparation.
\bibitem{sorella} S. Sorella, Phys. Rev. B {\bf 71}, 241103 (2005).
\bibitem{niu1985} Q. Niu, D.J. Thouless, and Y.-S. Wu, Phys. Rev. B {\bf 31}, 3372 (1985).
\bibitem{sheng2003} D.N. Sheng, Xin Wan, E.H. Rezayi, Kun Yang, R.N. Bhatt, and F.D.M. Haldane, Phys. Rev. Lett. {\bf 90}, 256802 (2003).
\bibitem{xwan} X. Wan, D.N. Sheng, E.H. Rezayi, Kun Yang, R.N. Bhatt, and F.D.M. Haldane, Phys. Rev. B {\bf 72}, 075325 (2005).
\bibitem{hafezi} M. Hafezi, A.S. Sorensen, M.D. Lukin, and E. Demler, Europhys. Lett. {\bf 81}, 1005 (2008)
\end{thebibliography}
\end{document}